# Manipulating moires by controlling heterostrain in van der Waals devices

Ian Sequeira[1†], Andrew Z. Barabas[1†], Aaron H Barajas-Aguilar[1], Michaela G Bacani[1], Naoto Nakatsuji[4], Mikito Koshino[4], Takashi Taniguichi[2], Kenji Watanabe[3], Javier D. Sanchez-Yamagishi[1*]

1. Department of Physics and Astronomy, University of California, Irvine, Irvine, CA, USA.
2. Research Center for Materials Nanoarchitectonics, National Institute for Materials Science, 1-1 Namiki, Tsukuba 305-0044, Japan
3. Research Center for Electronic and Optical Materials, National Institute for Materials Science, 1-1 Namiki, Tsukuba 305-0044, Japan
4. Department of Physics, Osaka University, Toyonaka, Osaka 560-0043, Japan

†these authors contributed equally to this work
*corresponding author

## Abstract:
Van der Waals (vdW) moires offer tunable superlattices that can strongly manipulate electronic properties. We demonstrate the in-situ manipulation of moire superlattices via heterostrain control in a vdW device. By straining a graphene layer relative to its hexagonal boron nitride substrate, we modify the shape and size of the moire. Our sliding-based technique achieves uniaxial heterostrain values exceeding 1%, resulting in distorted moires that are larger than those achievable without strain. The stretched moire is evident in transport measurements, resulting in shifted superlattice resistance peaks and Landau fans consistent with an enlarged superlattice unit cell. Electronic structure calculations reveal how heterostrain shrinks and distorts the moire Brillouin zone, resulting in a reduced electronic bandwidth as well as the appearance of highly anisotropic and quasi-1-dimensional Fermi surfaces. Our heterostrain control approach opens a wide parameter space of moire lattices to explore beyond what is possible by twist angle control alone.

## Introduction:
Moire materials offer highly-tunable superlattices with lengthscales that are unattainable in natural crystals. The extreme sensitivity of the electronic behaviors and correlated states to the moire structure motivates the precise control of moire patterns, especially in-situ manipulation techniques[1–4]. These efforts have primarily focused on controlling the relative twist angle, which determines the size of the moire pattern, but cannot alter its shape or symmetry.

Strain control offers an alternative route to manipulating moire patterns beyond what is possible via twist angle, where both the size and symmetry of the moire can be altered. To strongly distort the moire, it is necessary to strain adjacent layers by different amounts, which is referred to as heterostrain. The possibility of achieving large heterostrain is a unique feature of van der Waals (vdW) materials due to exceptionally low interlayer shear strengths[5]. When two lattices form a moire, the effects of heterostrain are amplified by a moire factor $\sim 1/(\delta-\varepsilon)$, where $\delta$ is the mismatch between the lattices and ε is the strain in one layer. This factor diverges as ε approaches δ, and has a directionality set by the strain direction. As a result, heterostrain can strongly distort the moire, changing its size and symmetries in a way not possible when straining both layers simultaneously (Figure 1a,b,c)[6,7]. Correspondingly, heterostrain can also strongly manipulate moire systems' electronic structure, especially in flatband systems such as twisted bilayer graphene[7–10].

To date, the experimental study of heterostrain has been limited by the challenge of controllably producing uniform and large heterostrain. Local probes such as scanning tunneling microscopy or optical measurements have been successful in measuring the effects of heterostrain on spectroscopic features[11–17], but transport measurements have been more limited due to the need for uniform heterostrain over larger areas. Current transport studies have relied on moire heterostrain introduced accidentally in the nanofabrication process[8,18]. Such strain profiles are generally small, and of uncontrolled direction and homogeneity. Progress has been made in controlling strain in vdW heterostructure devices by stretching or bending the underlying substrate[19–21], or by using deposited stressors[22–24]. For all approaches, an outstanding challenge is to achieve large and controlled heterostrain in a high-quality moire device.

## Results and Discussion:

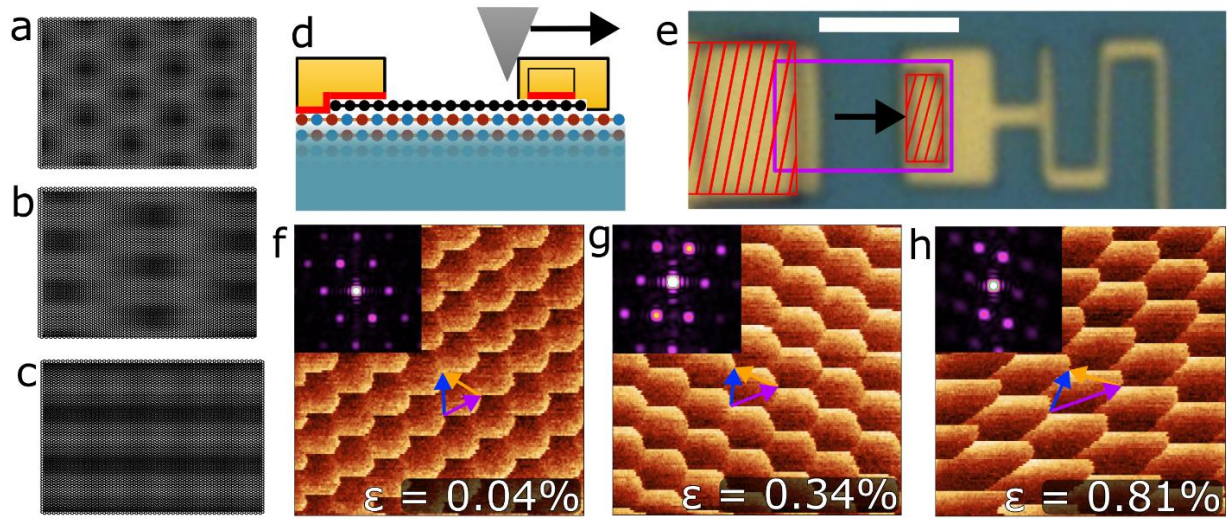

**Figure 1: Controlling heterostrain to manipulate vdW moires formed between graphene and hBN.**
**a-c:** Cartoon hexagonal lattices with a δ = 5% lattice mismatch. **a:** unstrained, **b:** 2.5% uniaxial heterostrain, **c:** 5% uniaxial heterostrain.
**d:** Side view schematic of a stretchable open-face g-hBN device. The graphene is stretched by pushing one electrode with an AFM tip (grey triangle). Red lines indicate areas where the metal is strongly adhered via O2-plasma treatment before deposition
**e:** Optical image of device B. Red dashed regions are O2-plasma treated. Purple box outlines the graphene channel. Scale bar is 4 μm

**f-h:** 100x100 nm CAFM images of g-hBN moires in device A. Inset shows FFT of the image. Fitting the wavevectors results in the indicated strain values ε (±3e-4 uncertainty) with strain angles relative to the +x-axis of +80±20°, +1±5°, +17±4°, and twist angles -0.12°, 0.02°, 0.02° for f,g,h, respectively. All parameters and uncertainties are listed in SI Table 1.

Here, we demonstrate the control of uniaxial heterostrain in graphene-hexagonal boron nitride (g-hBN) devices and study its effects on moire structure and transport properties. To achieve this, we have fabricated open-faced graphene devices on an hBN substrate where the electrodes are used to both measure transport and selectively strain the graphene (Figure 1d&e). The graphene is aligned to the hBN using a gold-based transfer method that deterministically produces large moire superlattices (see deterministic alignment of g-hBN samples in the Supporting Information (SI)).

Following our prior work on mechanically-reconfigurable vdW devices[25], we implement a moveable handle electrode that grips the graphene to slide and stretch it across the hBN. To selectively adhere the metal handle to the graphene only, we apply a light O2-plasma etch before depositing the Cr/Au handles. The graphene-handle bond is strong enough to strain graphene beyond its tensile strength and tear it (Figure S4a&b). We found the plasma pre-treatment to be critical, as thermally-evaporated metals such as Au, Cr, and Ni exhibit low sliding friction on graphene and cannot overcome the friction of graphene aligned to hBN.

The devices are designed with a fixed source electrode and a moveable drain electrode that can be displaced laterally by pushing with an atomic force microscope (AFM) tip (Figure 1d&e), inducing heterostrain in the gripped graphene layer. The moveable electrode has sufficient sliding friction to retain the graphene strain when the AFM tip is retracted. The end result is an open-face graphene device that can be progressively strained independent of the hBN substrate. The graphene strain can be directly observed both in the physical displacement of the graphene edges (measured in AFM), as well as in Raman shifts of the graphene G peak, both showing at least 0.6% strain can be fixed in the graphene layer (Figure S5a-d).

To image the effects of heterostrain on the g-hBN moire, we perform conductive AFM (CAFM) on the open graphene channels (Figure 1f-h). An unstrained graphene is shown in Figure 1f, where the moire pattern appears clearly in CAFM with a ~14 nm wavelength, the maximum possible for unstrained g-hBN. By stretching the graphene via the gold handle, we observe a large elongation of the moire along the strain direction, in line with the channel (Figure 1g&h). For the case in Figure 1h, the moire lattice vector most aligned to the strain direction is stretched by 84% to 25.7 nm.

To analyze the stretched moire structure, we extract the moire lattice vectors from the fast fourier transform (FFT) of the CAFM images. We find that a uniaxial strain model can fit most device regions to within 1% of the extracted wavevectors. By assuming pure uniaxial graphene heterostrain, we can extract local values for the uniaxial strain (ε), strain direction (φ), relative graphene-hBN twist angle (θ), as well as the global orientation of the layers with respect to the AFM image (See SI Uniaxial heterostrain model and fitting section).

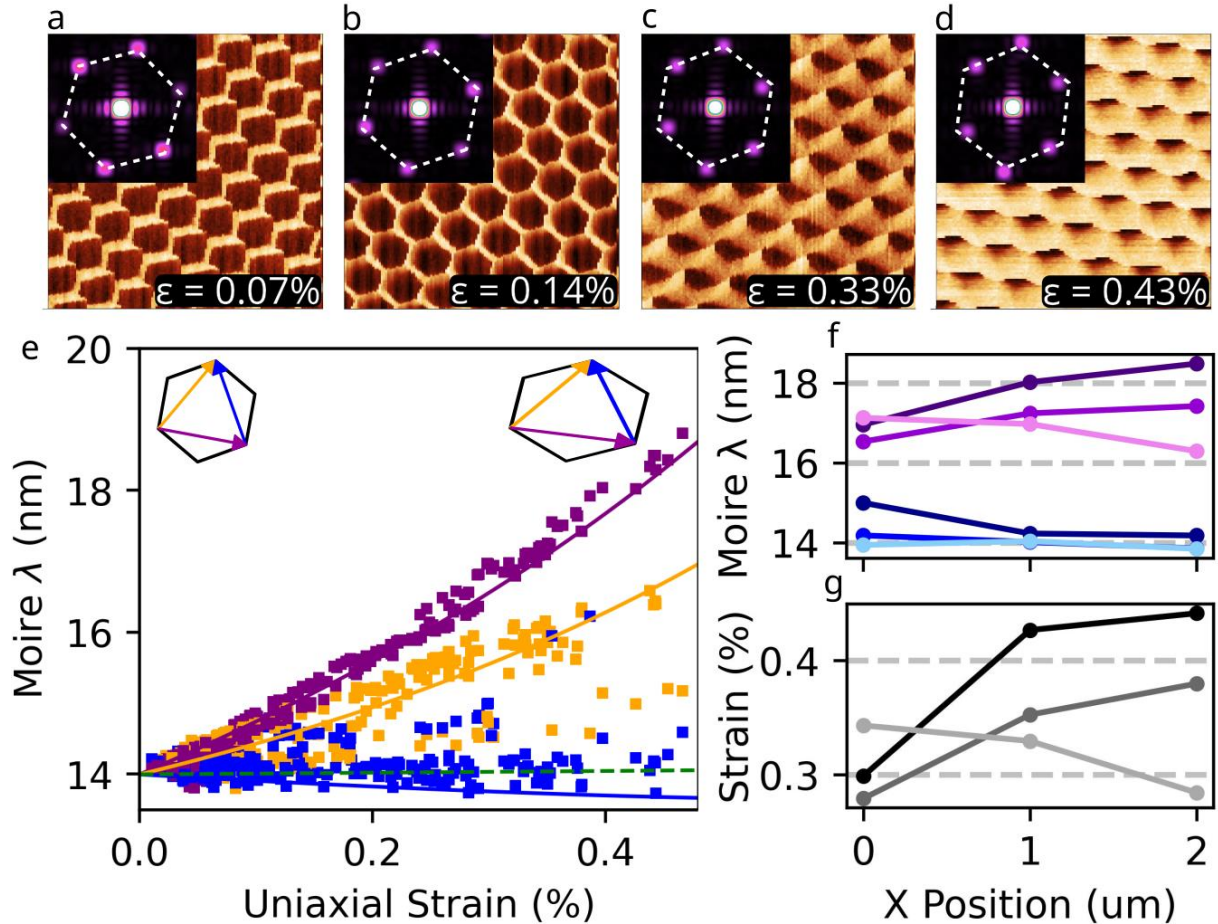

**Figure 2: Progressive uniaxial strain and spatial uniformity in device B.** All reported strain values have an average uncertainty of ±0.02% due to a 1% uncertainty in measuring the moire wavevector. All parameters and uncertainties are listed in SI Table 1.
**a-d:** Sequence of 100x100 nm CAFM images of graphene-hBN moires ordered from lower to higher strain (0.07% to 0.43%). Images are taken within ~200 nm of the same location, and the x-axis is aligned along the graphene channel. Insets: FFT of the real space images. White hexagon overlay is the same for each panel as a reference to demonstrate the evolution of the lattice wavevectors with strain.
**e:** Plot of measured moire wavelengths versus uniaxial graphene strain for ten strained states. Lines are theoretical moire wavelengths for a twist angle of 0.022° and strain angle of 18.279° with respect to the graphene lattice vector, that correspond to the average values for the dataset. Insets illustrate the moire unit cell orientation and stretch.
**f:** Moire wavelength versus position for the most strained state. Blue/purple shaded points correspond to the blue/purple vectors in Fig 2e, respectively. Different hues correspond to different rows along the devices separated by vertical distances of 1 um.
**g:** Graphene strain versus position for a single strained state. Each line is a row along the device.

To track the evolution of the moire patterns under strain, we incrementally stretch our samples and perform CAFM imaging in a grid of locations to characterize the homogeneity.

Figures 2a-d show a sample of these for device B, as it is progressively stretched horizontally. We measure this sample in ten strained states and plot the measured moire wavelengths versus the strain values extracted from our uniaxial strain model fits (Figure 2e). As the graphene is stretched, we observe that the moire lattice vectors are elongated depending on their orientation to the strain axis. The (purple) moire vector, nearly aligned with the strain axis, is stretched up to 34% relative to 14 nm. This corresponds to a 0.46% uniaxial strain of the graphene aligned 5±4° with respect to the AFM x-axis and the graphene channel. The (blue) moire vector that is nearly perpendicular to the strain axis shows little change. The theoretical moire wavelength for a 0° sample under uniaxial homostrain is plotted in green and shows weak dependence on strain. It is clear then that by heterostraining our samples we are able to achieve moire wavelengths that would be impossible using homostrain.

We compare the moire wavelength measurements to the expected values given by the average twist angle ($\theta = 0.022°$) and uniaxial strain angle ($\varphi = 5°$ with respect to the horizontal) extracted for all the points (solid lines in Figure 2e). A majority of the datapoints are clustered around the model lines, indicating the uniformity in the twist and strain angles in the sample. See SI figure S8 for an example of a stretched device with a non-uniform twist angle.

The stretched moires exhibit a consistent spatial variation across the channel, with the largest moire wavelengths and strains near the stretching electrode. This effect is clearly seen in the final strained state in Figure 2 f&g. We attribute this strain gradient to the friction of the aligned g-hBN domains, as well as local pinning from sample edges and disorder, which prevents an ideal elastic response of the stretched graphene. Nonetheless, in the final state, the device is strained across its entire 2.1 µm length with a max wavelength that varies from 16.3 to 18.8 nm.

The large heterostrain with micron-scale uniformity achieved with our approach offers a unique opportunity to study the electronic properties of stretched moires in transport. When graphene is aligned to hBN, superlattice-induced band gaps emerge at the edge of the reduced moire Brillouin zone (Figure 3a). In transport measurements, this results in new superlattice resistance peaks at densities n = ±4/(moire unit cell area), where the peak at hole doping is more prominent (Figure 3b). As the aligned graphene is stretched, we observe the superlattice resistance peak to progressively shift closer to the primary Dirac peak at charge neutrality. Compressing the channel shifts the peak in the opposite direction. This effect is consistent with heterostrain causing a substantial modification of the moire unit cell area. At the maximum strained state, the superlattice peak has moved 29% closer to the main Dirac peak, suggesting a corresponding increase in the moire unit cell area.

To further analyze the behavior of the superlattice resistance peak, we compare moire areas extracted from the resistance peak position to moire areas measured by CAFM. We plot these in Figure 3c for multiple devices in different strained states. Due to the applied heterostrain, many of the data points exceed the maximum area for unstrained, aligned g-hBN of $\sqrt{3}/2\ (14nm)^2 \sim 170\ nm^2$ (black diamond). The horizontal spread in CAFM areas is due to strain and twist angle inhomogeneity within each device. For small moires, there is agreement between the areas extracted from transport and CAFM. However, for stretched moires beyond ~14 nm, the moire area extracted from transport consistently exceeds the area measured in

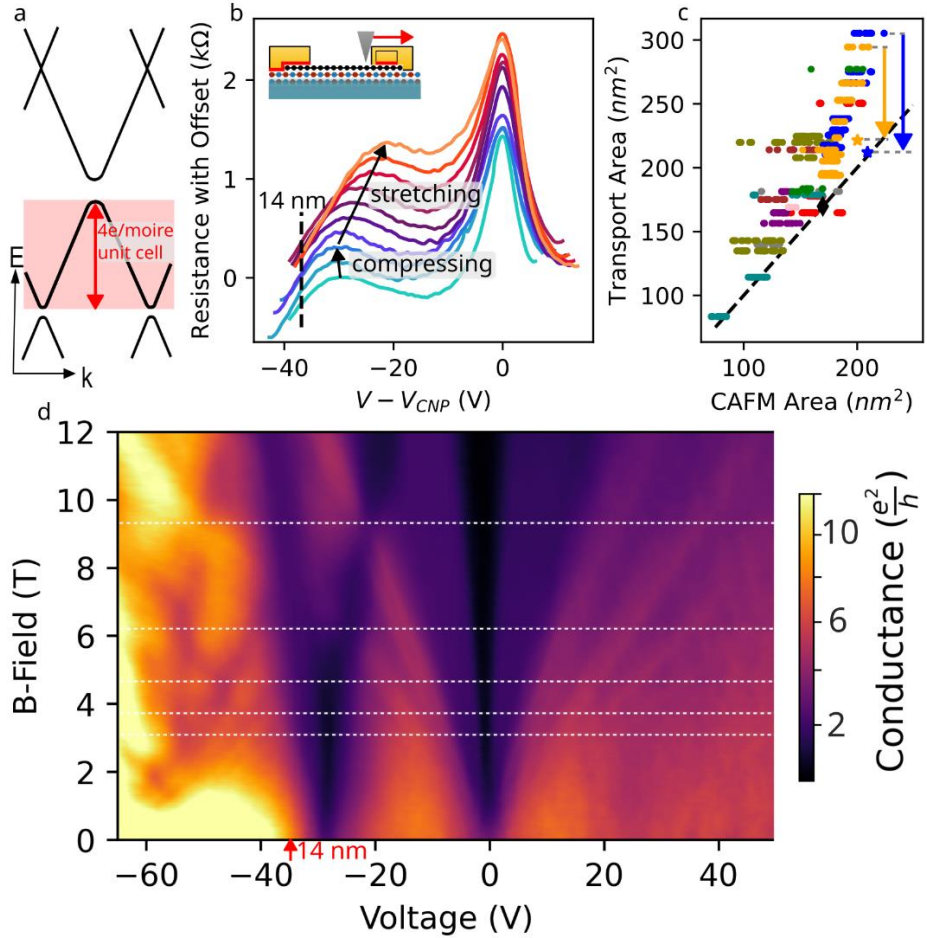

**Figure 3: Effects of uniaxial heterostrain on the transport properties of graphene-hBN devices.**

**a:** Cartoon band structure of aligned g-hBN at zero strain. Each moire subband corresponds to 4 electrons per moire unit cell area.

**b:** Room temperature gate sweeps of device B showing the superlattice peak shifting due to AFM manipulation. Sequential traces are offset for clarity. The measurements are made *in situ* to the AFM immediately after manipulating. The first two manipulations are compressive and the last seven are stretching motions. The dashed black line indicates the expected location of the superlattice peak for the largest unstrained moire, corresponding to a 14 nm wavelength. Inset: device manipulation schematic.

**c:** Moire unit cell area measured from CAFM images versus the area extracted from the superlattice resistance peak position via transport. Marker colors differentiate 10 devices. Circle markers correspond to transport data taken at ambient conditions. The two star markers correspond to transport data taken at 1.6K (plotted at the average CAFM area). Devices B/C are colors blue/orange, respectively. The black dashed line denotes equal CAFM and transport areas.

**d:** Magnetoconductance data of device C taken at 1.6K. Horizontal white lines overlaid correspond to a magnetic flux of 1/2,1/3,1/4,1/5,1/6 flux quanta per moire unit cell area ($221 \pm 3 nm^2$). A contact resistance value of 4.43 k$\Omega$ is subtracted to align the $\nu = 2$ plateau to 2 $e^2/h$.

CAFM. We ascribe this effect to the reduced electronic bandwidths of the stretched g-hBN devices. Similar effects have been observed in room temperature measurements of twisted bilayer graphene[18,26]. Indeed, cooling down devices B and C to 1.6 K removes the discrepancy between the transport and CAFM areas (see SI superlattice dirac peak temperature dependence section).

At cryogenic temperatures, applying a perpendicular magnetic field to devices with nearly-uniform heterostrain results in Landau fans emerging from the charge neutrality point and the hole-side superlattice peak (Figure 3d and Figure S11). Figure 3d shows magnetotransport data for device C, which has an average uniaxial strain of 0.30±0.04%, twist angle of 0.03±0.04°, and strain angle of 9±4° with respect to the graphene lattice vector, where the uncertainties are the standard deviation from the CAFM dataset. Fan features emerge from a sharp superlattice peak corresponding to a moire area of 221±3 nm$^2$, which is close to the average areas measured by CAFM at room temperature (200±10 nm$^2$). Due to the stretched moire unit cell size, the primary Landau fan is disrupted at lower magnetic fields than is typical for g-hBN devices[27,28]. These disruptions arise due to the collisions of the primary and superlattice Landau fans, with intersections occurring at B field values of 1 flux quanta per integer number of moire unit cells, $B = \varphi_0/(q \cdot A_m)$, where B is the magnetic field, $\varphi_0$ is the magnetic flux quanta, $A_m$ is the moire unit cell area, and q is an integer (horizontal lines in Figure 3d). We conclude that the heterostrain is sufficiently uniform to produce coherent superlattice modulation of the transport features consistent with a stretched moire unit cell.

Another characteristic of aligned g-hBN devices is the presence of an insulating state at charge neutrality due to the breaking of graphene inversion symmetry by the aligned hBN[28,29]. In the heterostrained devices, we only observe a weak insulating-like dependence at the charge neutrality point or at the superlattice point for our heterostrained devices. We ascribe this to strain and charge inhomogeneity in the device, where the latter is known to play an important role in obscuring insulating states in graphene devices due to edge doping effects[30,31].

To analyze the effects of heterostrain on the electronic structure of the g-hBN moire, we theoretically calculate the band structure using a continuum model that includes lattice relaxation (Figure 4, details in the SI theory section). The simplest effect of the uniaxial heterostrain is to shrink the moire Brillouin zone (BZ), reducing the carrier density per superlattice miniband. This results in a reduced bandwidth of the first hole miniband, which shrinks by 41.9% for 1% uniaxial strain applied along the graphene lattice vector (zero strain angle). Under the same strain conditions, the primary Dirac point gap grows by 22.7%, and the first hole superlattice gap decreases by 15.5%. Straining along different directions causes different changes to the gap sizes, but the primary and hole superlattice gaps remain open up to 1% strain, independent of strain angle.

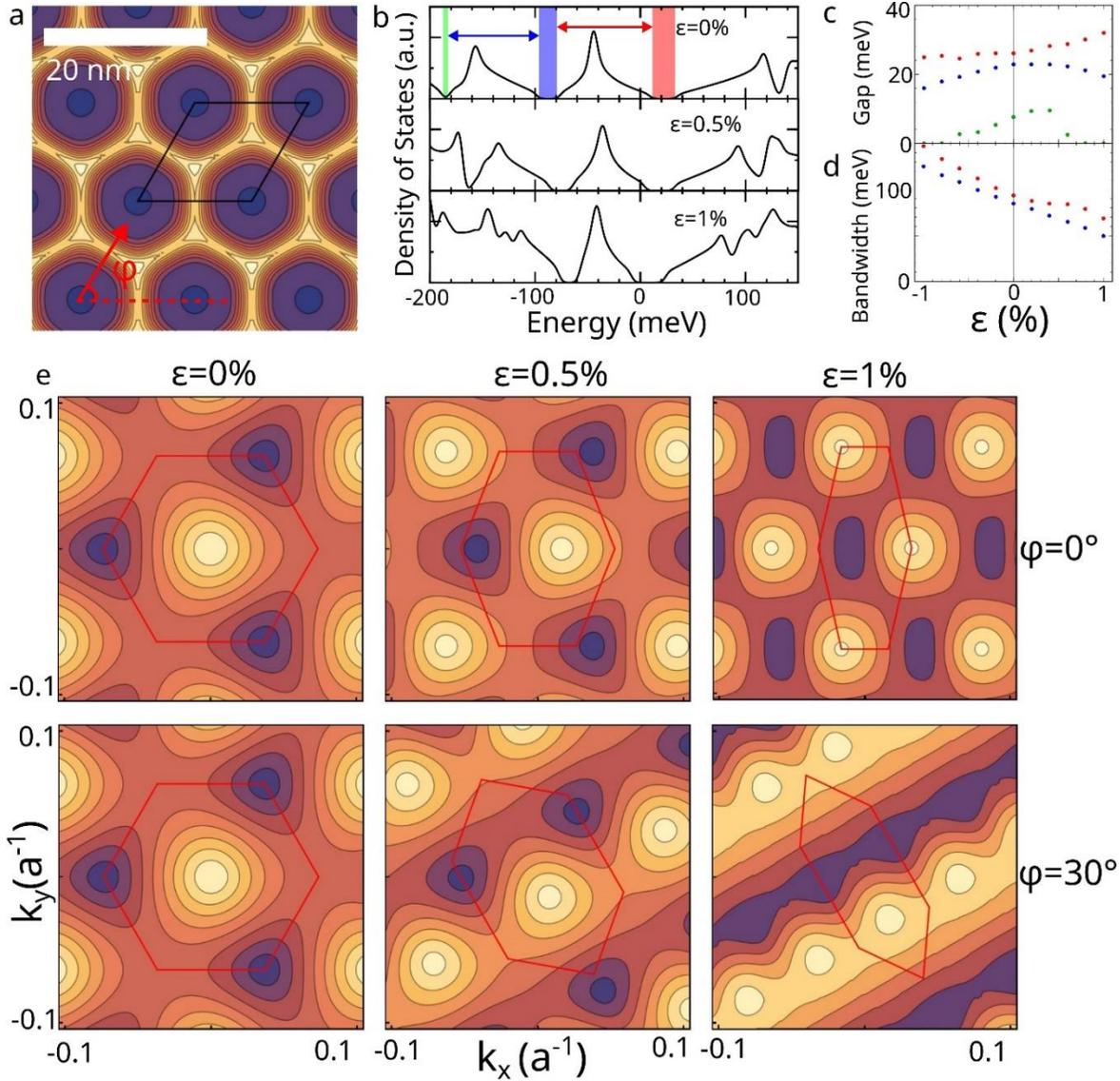

**Figure 4: Electronic structure of heterostrained graphene aligned to hBN with 0° twist angle**

**a**: Real space plot shows the local binding energy between the graphene and hBN lattices. Contour intervals are 0.2 eV/nm². The moire unit can strain angle orientation are illustrated.
**b**: Density of states plots for ε = 0%, 0.5%, and 1% at φ = 0°. Highlighted regions and arrows indicate gaps and bandwidths for **c** and **d**.
**c**: Primary, secondary, and tertiary gap size versus strain at φ = 0.
**d**. First and second hole miniband bandwidth versus strain at φ=0.
**e**. Band structure contour plots of the first hole miniband for ε=0%, 0.5%, and 1% at φ=0° and φ=30°. Wavevector units are expressed in terms of the graphene lattice constant, a. Red hexagons show the moire Brillouin zones. Contour intervals are 10 meV.

These basic theoretical results are in line with our experimental observations of the shifting of the superlattice resistance peak as strain increases the size of the moire unit cell. Because both the primary and hole-side superlattice gaps remain open as the graphene is stretched, the carrier density of the hole miniband remains at 4 electrons per superlattice unit cell, which is consistent with the match between our cryogenic transport measurements of the superlattice hole peak and the CAFM-extracted moire area. By contrast, the electron-side superlattice band edge remains gap-less with applied strain, consistent with our lack of observation of an electron-side superlattice resistance peak. Moreover, the decrease in bandwidth with uniaxial strain is consistent with the discrepancy with the moire area extracted from room temperature measurements as compared with CAFM and cryogenic measurements. At room temperature, thermal energy is a significant fraction of the bandwidth of the first hole miniband, with increased thermal excitation expected as the bandwidth is reduced with 1% heterostrain from 85 meV to 49 meV for zero strain angle (Figure 4).

Heterostrain introduces strong anisotropy in the graphene electronic structure, where the Fermi surface near the hole superlattice gap becomes stretched and can even become quasi one-dimensional due to the heterostrain. For zero strain angle, the anisotropy is apparent in the effective mass at the bottom of the hole miniband, where $m_y/m_x = 3.63$, corresponding to the ratio of the effective mass in the direction of the strain versus perpendicular for 1% uniaxial strain and zero strain angle. This is contrasted with the Fermi surfaces around the primary Dirac point, which are less sensitive to the distorted moire and remain effectively isotropic under the effects of the 1% graphene strain. For 30° strain angle, open Fermi surfaces dominate the electronic band structure for large heterostrain (Figure 4e). These open Fermi surfaces are strongly anisotropic and quasi-one-dimensional, where the Fermi velocity is limited to a range of directions along a preferred propagation direction.

# Conclusions:

In summary, we present a method to controllably introduce large heterostrain in a g-hBN device. The induced heterostrain is sufficiently homogenous to observe a modified electronic structure corresponding to the stretched moire pattern. These overall observations are in agreement with the reduced bandwidth and BZ size observed in theoretical calculations, which also predict that heterostrain should introduce anisotropy in the effective masses and open Fermi surfaces. In future studies, multi-terminal devices with different strain angles will be able to probe the effects of the anisotropic electronic structure. We note that even larger strain values are achievable locally using this technique, where we have observed local moire patterns ~70 nm long correspond to ~1.4% level strains (See SI Figure S9).

For this work, we used an AFM tip to set the strain in our devices, but we envision that other actuation methods such as bending, piezoelectric actuation, or thermal contraction can be used to achieve in-situ manipulation, even at cryogenic temperatures. Moreover, our methods should apply to other types of crystals supported on hBN, especially twisted graphene and transition metal dichalcogenide multilayers, where strain control can be used to deterministically break the symmetry of moire patterns, manipulate the structure of pseudomagnetic fields and electronic flatbands, or alternatively be used as a method to reduce strain in a device unintentionally introduced by nanofabrication.

# Supporting Information

Additional experimental details, materials, methods, and theoretical model (pdf)

## Acknowledgements

We acknowledge the use of facilities and instrumentation at the Integrated Nanosystems Research Facility (INRF), in the Samueli School of Engineering at the University of California Irvine, and at the UC Irvine Materials Research Institute (IMRI), which is supported in part by the National Science Foundation through the UC Irvine Materials Research Science and Engineering Center (DMR-2011967). We thank L. Jauregui, S. Nam, D. Goldhaber-Gordon, and A.F. Young for productive discussions, as well as the technical assistance of Q. Lin, R. Chang, M. Kebali, and J. Hes. We would also like to thank D. Velarde, F.A. Bonilla, and S. Kaemmer at Park Systems for their suggestions and technical assistance.

## Funding

This work was partially supported by the National Science Foundation Materials Research Science and Engineering Center program through the UC Irvine Center for Complex and Active Materials DMR-2011967 and the National Science Foundation Career Award DMR-2046849. I.S. acknowledges fellowship support from the UCI Eddleman Quantum Institute. K.W. and T.T. acknowledge support from the JSPS KAKENHI (Grant Numbers 21H05233 and 23H02052) and World Premier International Research Center Initiative (WPI), MEXT, Japan. N.N and M.K acknowledge support from the JSPS KAKENHI (Grants numbers JP21H05236, JP21H05232) and by JST CREST (Grant number JPMJCR20T3), Japan.

# Supporting Information: Manipulating moires by controlling heterostrain in van der Waals devices

Ian Sequeira[1†], Andrew Z. Barabas[1†], Aaron H Barajas-Aguilar[1], Michaela G Bacani[1], Naoto Nakatsuji[4], Mikito Koshino[4], Takashi Taniguichi[2], Kenji Watanabe[3], Javier D. Sanchez-Yamagishi[1*]

1. Department of Physics and Astronomy, University of California, Irvine, Irvine, CA, USA.
2. Research Center for Materials Nanoarchitectonics, National Institute for Materials Science, 1-1 Namiki, Tsukuba 305-0044, Japan
3. Research Center for Electronic and Optical Materials, National Institute for Materials Science, 1-1 Namiki, Tsukuba 305-0044, Japan
4. Department of Physics, Osaka University, Toyonaka, Osaka 560-0043, Japan

†these authors contributed equally to this work
*corresponding author

## Sample Fabrication

### Deterministic alignment of graphene-hBN samples

To maximize the impact of heterostrain on the size of the g-hBN moire pattern, the graphene must be aligned to the hBN, with the twist angle as close to zero degrees as possible. To reliably fabricate such aligned open-face g-hBN stacks, we developed a gold-based transfer technique and deterministically rotate the graphene into alignment using AFM nanomanipulation.

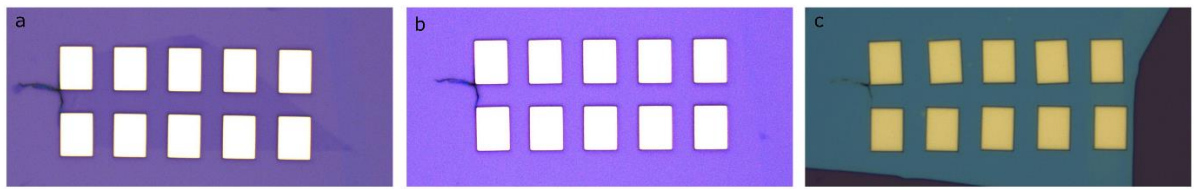

**Figure S1: Optical images of gold based transfers**
**a:** 200 nm thick gold shapes transferred onto a graphene flake. Shapes are 8 μm x 6 μm
**b:** After oxygen plasma etching to etch away exposed graphene
**c:** Etched gold-graphene stack transferred onto bottom hBN flake and ready to be aligned by performing AFM manipulations

We begin by lithographically defining rectangular gold shapes (4 μm x 8 μm x 200 nm) on SiO$_2$ and picking them up with a PC stamp. To pick up the gold shapes, we typically heat to approximately 150C, so the PC melts and completely encases the gold shapes, then cool to room temperature and pick up slowly to reduce strain in the PC film. Next, we transfer the gold shapes onto graphene and perform an O2 plasma etch to remove the exposed graphene (Figure S1 a&b). At this stage, the gold-graphene stack is picked up with another PC stamp and transferred onto a bottom hBN substrate (Figure S1c).

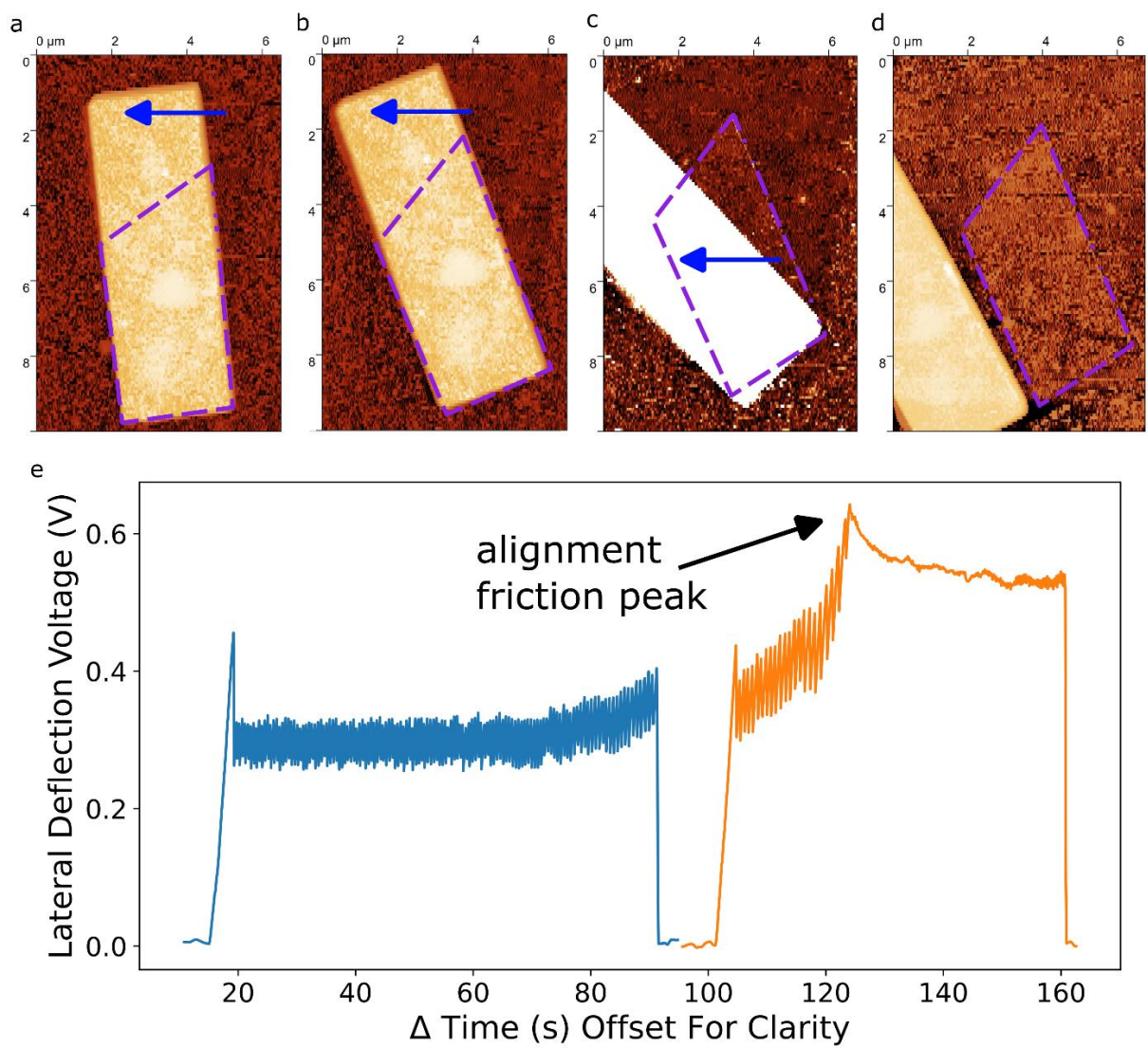

**Figure S2: Deterministic alignment of graphene to hBN substrate**
**a-d:** Sequence of AFM topography images of gold-graphene on hBN during the AFM manipulations to rotate the graphene into alignment, purple dashed lines outline the graphene. Blue arrows indicate positions where the gold shape is pushed with the AFM tip. **d** is the final result after sliding the gold off of the aligned g-hBN
**e:** Lateral AFM deflection voltage versus time for the manipulations done in **a** (blue) and **b** (orange). As the graphene gets rotated into alignment with the hBN, there is a peak in the voltage signal and the gold slips off.

To rotate the graphene, the Au is pushed laterally at one end with an AFM tip (Figure S2 a&b). This rotates the Au, and the friction between the Au and graphene rotates the graphene along with it (see AFM manipulation technique section). Near 0°, the g-hBN friction peaks due to increased atomic registry [1–4] (Figure S2e). Ideally this friction exceeds the Au-graphene friction, causing the graphene to lock to the hBN and the Au to slide off of the graphene and leaving behind a well-aligned, clean, open-face g-hBN stack (Figure S2 c&d). In practice though, the g-hBN friction is not always enough to release the graphene from the Au by rotating. Often we find that it is necessary to halt the rotation at the aligned g-hBN friction peak, then translate the Au off of the graphene.

A common failure when rotating or translating the Au off, is for the graphene to be torn from its edges. We speculate this is due to the edges of the graphene having bonded to the gold. Nevertheless, we found this process proved to be significantly more reliable in producing well-aligned (<0.5°), open-face devices as compared to simply aligning straight edges of the two flakes during the transfer process. Lastly, to further clean the surfaces and aid in further alignment we vacuum anneal the samples at 350C for 30 minutes.

## Flexible electrode and handle contacts

We perform 2 steps of electron beam lithography (EBL) and metal vapor deposition, with the first step primarily to mechanically contact the graphene and the second to electrically contact.

Using EBL, we define a large contact on one side of the graphene which overlaps with the hBN. This will be the side of the graphene which is stationary and pinned to the hBN. On the other side of the graphene strip we define a rectangle in the interior of the graphene, so that it does not overlap with the hBN at all. This will be the handle which we will use to apply forces to the graphene. We then etch these features with an O2 plasma. The etch is performed with a PE-25 plasma cleaner on low power (~10W), with 15 SCCM of $O_2$ at 250 mTorr for 20s. We then deposit 5/95 nm of Cr/Au.

For the 2nd metal contact step we overlap the previous contacts in order to improve electrical contact to the graphene. On the handle side of the strip we write the electrode in a thin, serpentine shape in order to increase compliance and decrease friction for AFM manipulation. This deposition is also 5/95 nm Cr/Au.

## AFM tip cleaning

Post device processing, the graphene will have substantial polymer residue on it which can interfere with CAFM imaging and is also a major source of disorder, broadening transport features. To clean the surface, we tip clean the sample with a Budget Sensors Multi75-G using a setpoint force between 5-15 nN. Before tip cleaning, superlattice peaks are typically not visible. After tip cleaning, the Dirac peak becomes sharper and the superlattice peaks are clearly resolvable.

## AFM cutting

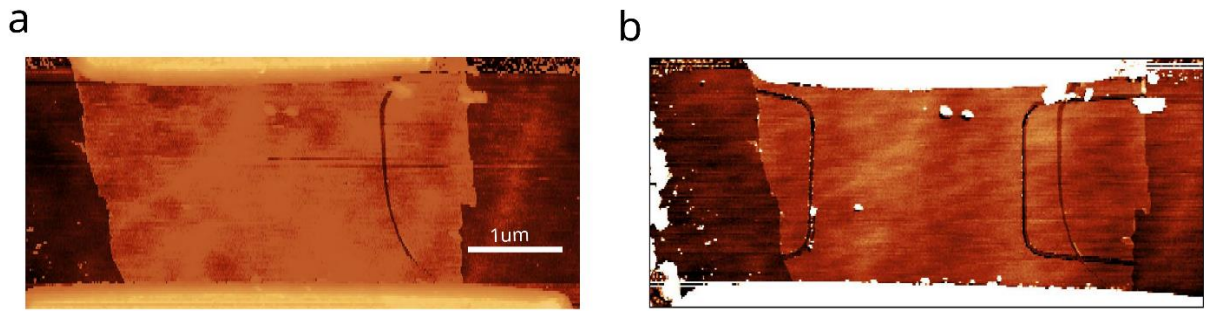

**Figure S3: Device E before (a) and after (b) AFM cutting**

We find that strain-induced tears tend to originate from nicks and sharp corners in the graphene. We remove these features by AFM cutting the graphene into a regular channel with rounded corners (~0.5 μm radius of curvature). We use AC local anodic oxidation[5] to cut the graphene in our Park NX10 AFM. We use a Nanoworld ARROW-EFM tip, 100 nN setpoint force, and 100 kHz, 20 $V_{pp}$ tip bias.

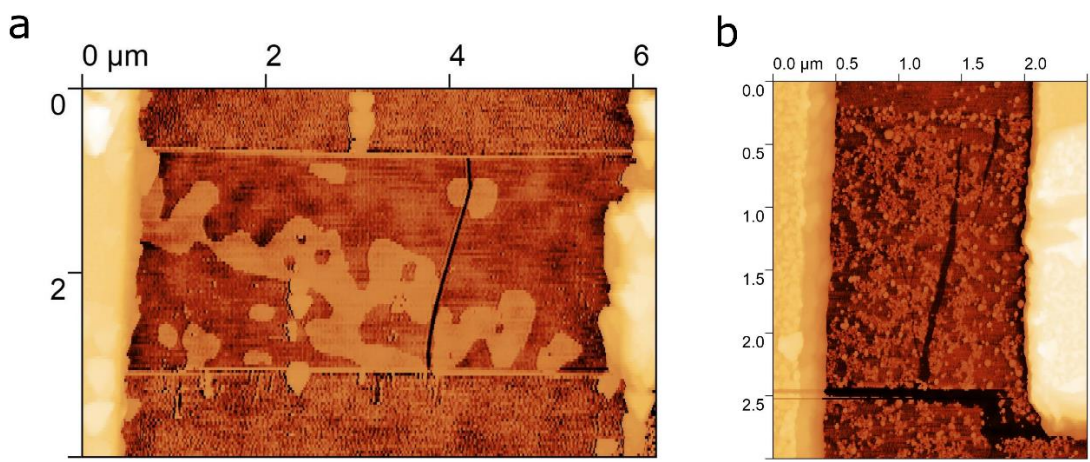

**Figure S4: Torn g-hBN devices.**
**a,b:** Graphene strips tore due to being stretched. Close inspection of (**a**) showed it tore at a nick in the edge.

## Current Annealing

Large contact resistances were observed in many of the devices. To decrease the contact resistance, we current anneal by voltage biasing while measuring the current with a Keithley 2400 source measuring unit to perform IV measurements. Typically the samples were current annealed up to 0.5 mA/um [6]. After current annealing we observed significant reduction in the contact resistance and hysteresis.

## Transport measurements

Two probe transport measurements were performed via current biasing with a 1 MΩ resistor at a frequency of 17 Hz and using an SRS830 lock-in amplifier to measure the voltage drop. At room temperature, current biases of 10 µA or 100 µA were used in order. At cryogenic temperatures, current biases of 100 µA or 1 µA were used.

For the g-hBN moire, the first hole doped superlattice gap is larger than its counterpart on the electron doped side[7,8]. At room temperature, thermal broadening and strain inhomogeneity often obscure the superlattice resistance peak on the electron-doped side of the Dirac point. For this reason, and to minimize hysteresis which can occur at ambient conditions in a silicon-backgated sample, we typically constrain our room temperature gate sweeps to the hole-doped side.

The open face g-hBN samples are fabricated on a highly doped silicon substrate with a 285nm SiO2 thickness. To calculate the density of the superlattice peak position and infer the moire unit cell area, a parallel plate capacitor model with a SiO2 dielectric in series with a Xnm hBN dielectric is used assuming dielectric constants 3.9 and 3.4 respectively. The uncertainty in the moire unit cell area extracted from the superlattice peak position is quoted as half the gate voltage step size.

## AFM manipulation technique

We stretch our devices using the Lithography mode of our Park NX10 AFM with the 1200 kHz tip on the MikroMasch 4XC-NN probe. We choose this very stiff tip (100 N/m) for manipulations because softer tips often exceed the 10 V limit of our AFM photodiode when manipulating higher friction objects and stiffer tips exhibit less backlash when transitioning from static to kinetic friction. We first measure the height of the gold handle surface. For the manipulation, the Z piezo is set to ~150 nm below the top surface of the gold handle, and the tip moves at 1 nm/s. During motions, we monitor and record both the AFM lateral deflection signal and the device resistance in order to better understand when the device begins moving. We have observed that friction for the initial motion is higher than for subsequent ones. This usually results in a large initial displacement on the order of 50 nm, which for short devices, is enough to immediately tear the graphene. For this reason, our first motion is to compress the device, so the initial large displacement does not rip the graphene.

## Unaligned graphene-hBN strain Raman signal

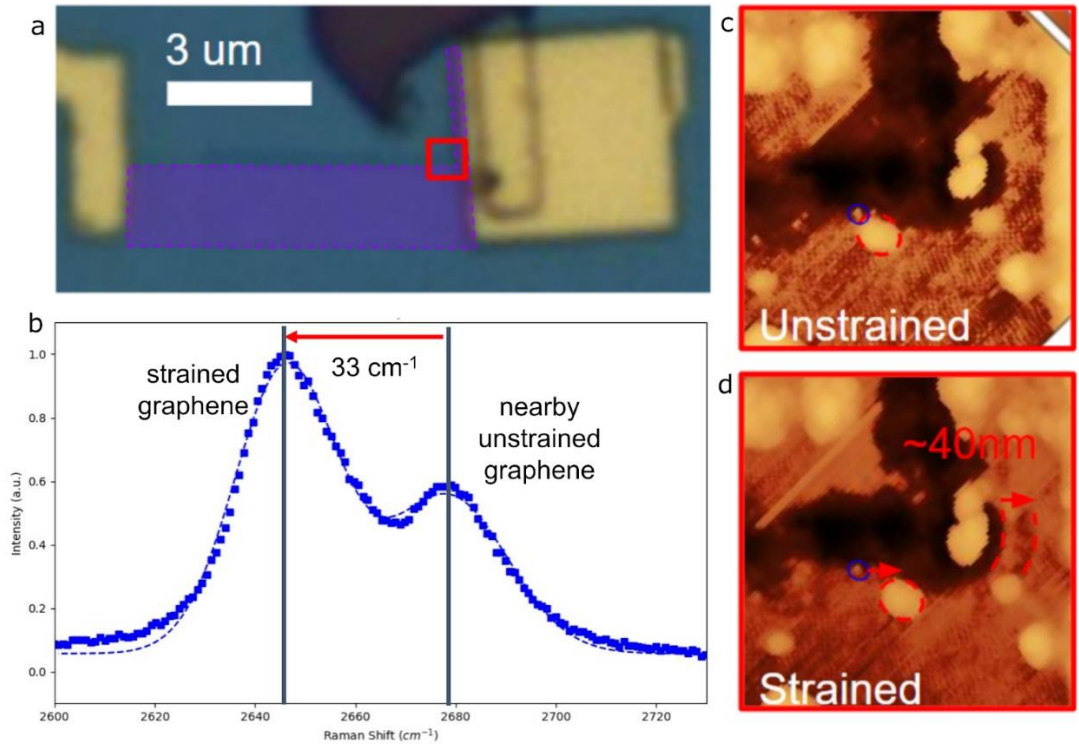

**Figure S5: Stretching and Raman of unaligned g-hBN**
**a:** Optical image of unaligned g-hBN device. Graphene strip highlighted in purple
**b**: Raman signal from strained graphene. Spot size also includes an adjacent, unstrained graphene strip. 33 cm$^{-1}$ shift corresponds to 0.61% graphene strain[9]
**c:** AFM image of graphene corner before motion
**d:** AFM image of graphene corner after motion, showing ~40 nm displacement relative to hBN substrate. 40 nm displacement for a ~6.78 μm long strip corresponds to 0.59% strain. This agrees well with the Raman signal

## Conductive AFM

      Our conductive AFM (CAFM) images are taken using the internal lock-in of our Park NX10 AFM. We apply an AC voltage to the sample and amplify the tip current signal using a FEMTO current amplifier. We use an ARROW-EFM tip with 50 nN setpoint force. Other tips, like the ARROW-NCPt work as well, but we prefer the ARROW-EFM because its lower spring constant makes it better suited for contact mode. These tips have a PtIr coating, and we have found that we are able to image moires with them for longer than TiIr coated tips, and much longer than Au or Pt coated tips.

      We take several steps to account for the drift in our conductive AFM images and reduce its effects. First, we minimize the time per scan by using a relatively high scan rate of 15 Hz, taking small scale images (100x100 nm), and only taking 128 lines per scan. At each moire position, we sequentially record a forward and backward scan image and extract a thermal drift velocity by comparing the differences in the extracted moire lattice vectors. With this information

we can correct for the effects of the drift in each image. Additional details can be found in the SI AFM drift correction section.

## Moire FFT strain fitting

To determine the strain parameters from the g-hBN moire patterns, we adapted a fitting approach from Tran et al[10]. The code generates graphene and hBN lattice vectors, transforms them via a global rotation, a relative g-hBN rotation, and a uniaxial strain. From these transformed lattice vectors we can generate the moire lattice vectors, and iteratively adjust rotation angles and strain to match the moire lattice measured from CAFM images. We use $a_G$ = 0.246 nm and $a_{hBN}$ = 0.2504 nm for the graphene and boron nitride lattice constants, respectively, and a graphene Poisson ratio = 0.16[11,12].

## Error Analysis

To estimate the uncertainty for the extracted device parameters, we numerically invert functions for calculating the twist angle, strain and strain angle parameters for a given set of moire wavevectors. We assume a 1% error in the moire wavevectors, and report the max and min range of the resulting calculated parameters as the uncertainty values in the main text. We estimate a 1% error in the FFT position from the reproducibility of repeated CAFM moire imaging in a single position (See SI Figure S6: AFM moire wavelength reproducibility).

## AFM drift correction

To correct for thermal drift in our AFM images of moire patterns, we record each AFM image twice with both forward and reversed slow scan axis. Due to the different scan directions, the images will be distorted differently by the drift. By comparing the two images and assuming a constant drift velocity and constant tip velocity we can determine the drift and correct for its effects. We model the drift in our AFM images as follows: For a square image with dimension *d* taken at scan rate *f*, the time elapsed between 2 points in a scan is:

$$\Delta t = \frac{\Delta x}{d}\frac{1}{2f} + \frac{\Delta y}{d}\frac{l}{f}\frac{1}{f} \quad \text{(A1)}$$

*d* = Image dimension
*f* = scan rate
*l* = number of lines in the slow direction
*x* = x position or size
*y* = y position or size

Given 2 moire lattice vectors **a** and **b**, one from an image taken top-to-bottom and the other taken bottom-to-top, the drift velocity, *v*, can be calculated, assuming it is constant for the duration of the 2 scans. Both vectors are an attempt to measure the sample's true moire lattice vector, but are distorted due to the drift, and thus each differ from the true vector by an offset Δ**a** and Δ**b** (the offsets are different from each other because **a** and **b** are different lengths, and thus have different elapsed times, meaning different displacements).

$$\vec{a} + \Delta\vec{a} = \vec{b} + \Delta\vec{b} \quad \text{(A2)}$$

The drift velocity can then be calculated from this offset and from the time it took to measure **a**. We know how much time elapsed between any two points in the AFM image, Δt, from above. And the distance the AFM has drifted in that time is Δ**a.**

$$\vec{v} = \Delta\vec{a}\left(\frac{a_x}{d}\frac{1}{2f} + \frac{l}{d}\frac{a_y}{f}\frac{1}{f}\right)^{-1} \quad (A3)$$

The same is also true for **b**.

$$\vec{v} = \Delta\vec{b}\left(\frac{b_x}{d}\frac{1}{2f} + \frac{l}{d}\frac{b_y}{f}\frac{1}{f}\right)^{-1} \quad (A4)$$

A2, A3, and A4 form a system of equations whose solutions are the drift velocity and offsets. We get 2 systems of equations from each pair of images, one from the **a₁** vector and one from **a₂**.

Performing this analysis on all of the moire AFM data we present in this paper, we find a relatively low drift: in the y direction 0.08±0.16 nm/s and in the x direction 0.00±0.04 nm/s corresponding to a 1±2 nm drift over the course of a single scan. There is significant spread which we attribute to having the AFM piezo scanners warmed up to different extents between measurements and other thermal effects.

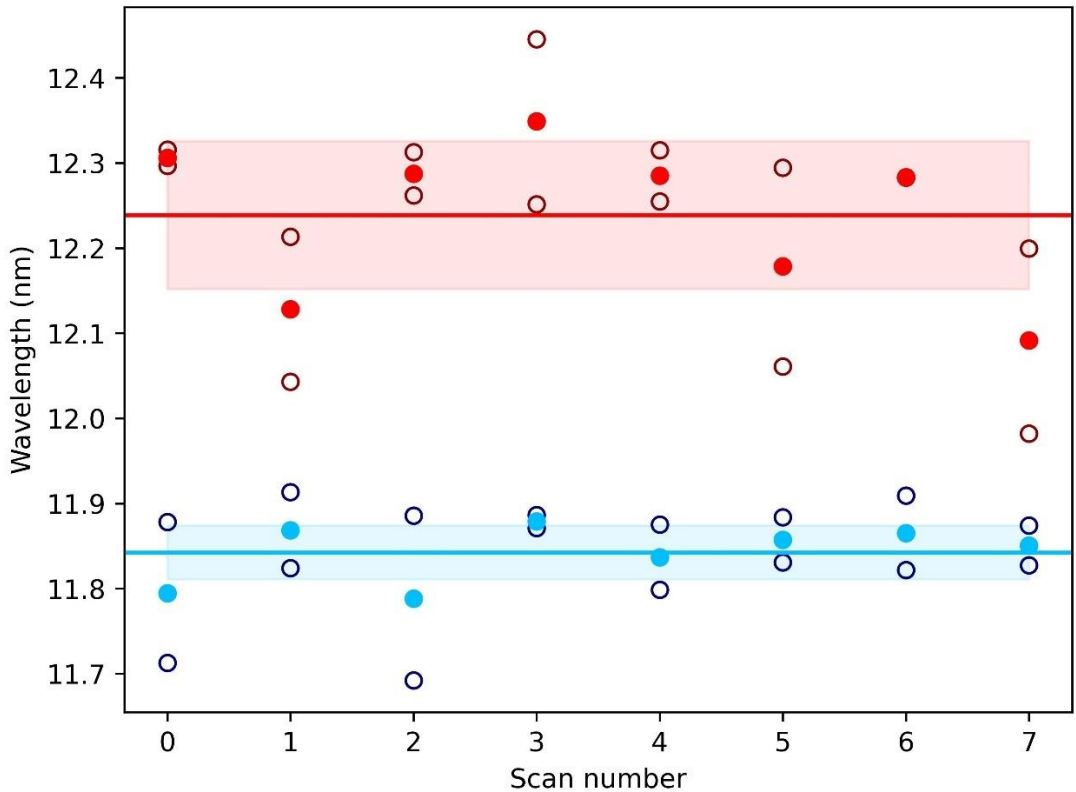

**Figure S6: AFM moire wavelength reproducibility**
Wavelength values were extracted from sequential AFM images at the same position over the course of 8 minutes.

Each scan number represents data taken from 2 AFM images, one forward and the other reversed. The hollow data points are the 2 raw values, which have not been corrected for drift, and the solid data points are the drift corrected values. Blue and red points denote different moire lattice vectors. The mean and standard deviation of the corrected values are shown with the horizontal lines and shaded areas, which are 12.24 ± 0.09 nm and 11.84 ± 0.03 nm. This corresponds to a 0.7% and, 0.3% spread in the moire wavelengths, respectively.

Even scan numbers consist of one image with the slow direction upwards and one downward; odd scan numbers have one rightwards and one leftwards. This explains the alternating spread in uncorrected wavelength.

We calibrate our AFM's XY scanner using an etched $SiO_2$/Si calibration standard provided by Park Systems by taking a 50x50 um image of the 5 um pitch standard.

## Uniaxial heterostrain model and fitting

We transform the real space graphene unit vectors with a rotated strain matrix:

$$(I + R(\varphi) \, S \, R^T(\varphi)) \, \vec{G}$$

Which is then rotated by the graphene-hBN twist angle, θ:

$$R(\theta)(I + R(\varphi) \, S \, R^T(\varphi)) \, \vec{G}$$

This result and the hBN unit vectors are rotated by a global twist angle. Their reciprocal lattice vectors are calculated. And the moire reciprocal lattice vectors are calculated by subtracting the rotated hBN reciprocal lattice vector from the stretched/rotated graphene reciprocal lattice vectors.

I is the identity matrix, and R and S are defined as follows:

$$R(\theta) = \begin{pmatrix} \cos(\theta) & -\sin(\theta) \\ \sin(\theta) & \cos(\theta) \end{pmatrix}; \quad S = \begin{pmatrix} \epsilon & 0 \\ 0 & -0.16\epsilon \end{pmatrix}$$

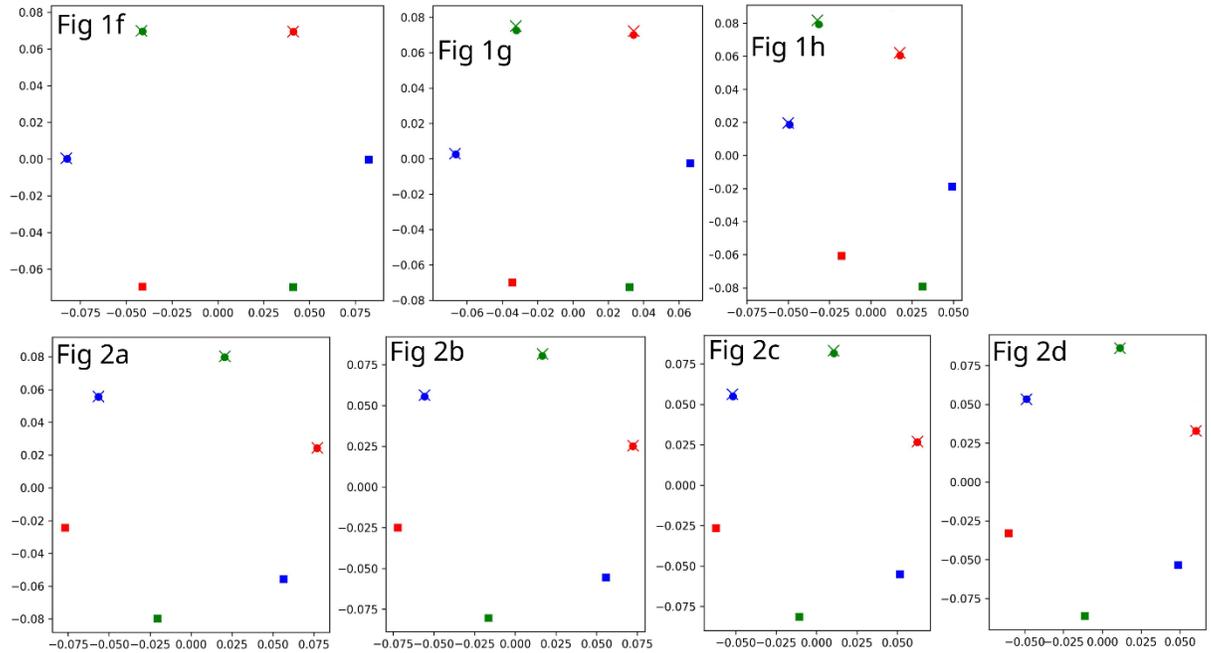

**Fig S7: Moire peak fitting**
Moire fits from the CAFM images of main text figures. Solid circles are the peak positions measured from the FFTs of CAFM moire images; solid squares are the same peaks inverted; crosses are the fitted peak positions using a uniaxial strain model to fit the measured peaks. The largest discrepancies in fitting result from the images which appear to require some biaxial strain to fit, such as Fig 1f&g. 85% of our moires can be fit to within 1% of the measured FFT peaks, and >99% can be fit to within 5%.

|  | Uniaxial strain $\varepsilon$ (%) | | Graphene-hBN twist $\theta$ (°) | | Strain angle wrt graphene $\varphi_G$ (°) | | Global twist $\theta_{global}$ (°) | | Strain angle wrt horizontal $\varphi$ (°) | |
|---|---|---|---|---|---|---|---|---|---|---|
| **Fig 1f** | 0.04 | +0.02 -0.02 | -0.1 | +0.1 -0.1 | 50 | +20 -20 | 30 | +7 -9 | 80 | +20 -20 |
| **Fig 1g** | 0.34 | +0.02 -0.02 | 0.0 | +0.1 -0.1 | -26 | +4 -5 | 27 | +8 -7 | 1 | +5 -4 |
| **Fig 1h** | 0.81 | +0.02 -0.01 | 0.0 | +0.1 -0.1 | -7 | +4 -5 | 24 | +9 -7 | 17 | +4 -3 |
| **Fig 2a** | 0.07 | +0.03 -0.02 | 0.0 | +0.1 -0.2 | -12 | +10 -10 | -15 | +9 -7 | -30 | +10 -10 |
| **Fig 2b** | 0.14 | +0.02 -0.02 | 0.0 | +0.1 -0.1 | 11 | +6 -6 | -13 | +9 -7 | -2 | +7 -7 |
| **Fig 2c** | 0.33 | +0.02 | 0.0 | +0.1 | 16 | +4 | -11 | +8 | 5 | +5 |

|  |  |  | -0.02 |  | -0.1 |  | -4 |  | -7 |  | -4 |
|---|---|---|---|---|---|---|---|---|---|---|---|
| **Fig 2d** | 0.43 | +0.02 -0.02 | -0.18 | +0.06 -0.05 | 0 | +2 -2 | 2 | +3 -4 | 2 | +2 -2 |
| **Fig 2e** | 0.05 | +0.03 -0.02 | 0.02 | +0.1 -0.1 | 18 | +10 -20 | -13 | +8 -7 | 5 | +20 -10 |
| **Fig 2e** | 0.10 | +0.02 -0.02 | 0.02 | +0.1 -0.1 | 18 | +8 -8 | -13 | +8 -6 | 5 | +8 -7 |
| **Fig 2e** | 0.20 | +0.03 -0.02 | 0.02 | +0.1 -0.2 | 18 | +5 -6 | -13 | +9 -6 | 5 | +6 -5 |
| **Fig 2e** | 0.30 | +0.02 -0.02 | 0.02 | +0.1 -0.1 | 18 | +4 -5 | -13 | +9 -7 | 5 | +5 -4 |
| **Fig 2e** | 0.40 | +0.02 -0.01 | 0.02 | +0.1 -0.1 | 18 | +4 -4 | -13 | +8 -6 | 5 | +4 -3 |

**Table 1: Parameters and uncertainties for main text figures**
The uncertainties correspond to a 1% error in the moire FFT peak position. The rows for Fig 2e each use the average twist angle, strain angle, and global rotation angle for the Fig 2e data with several uniaxial strain values in order to estimate the uncertainty throughout the data range.

## CAFM and strain data from other devices

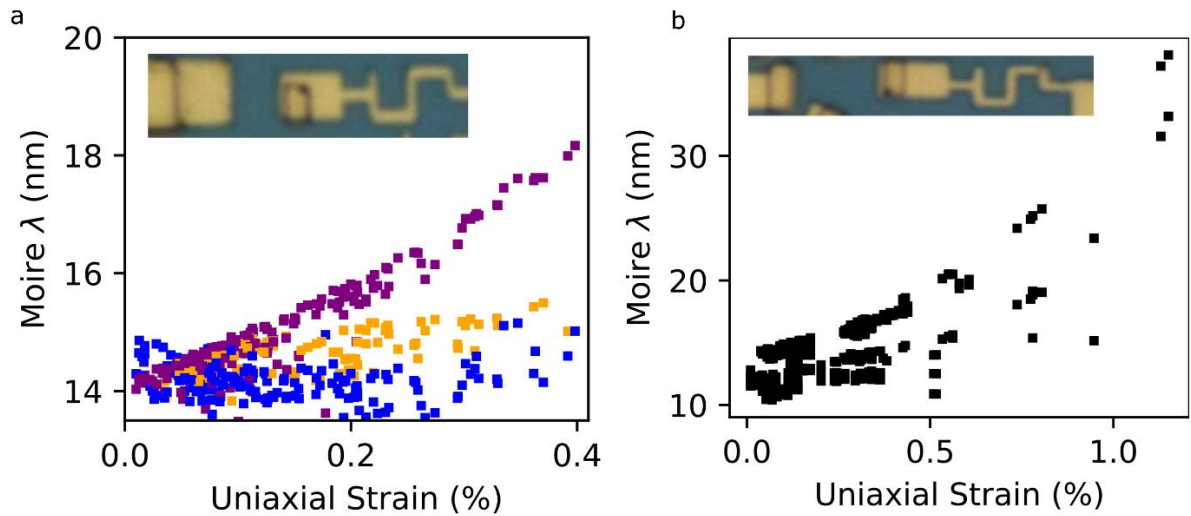

**Figure S8: Moire wavelength versus strain for Device C (left) and Device A (right)**
Data for Device A are not differentiated by color because the twist inhomogeneity and high strain make different moire vectors difficult to distinguish.

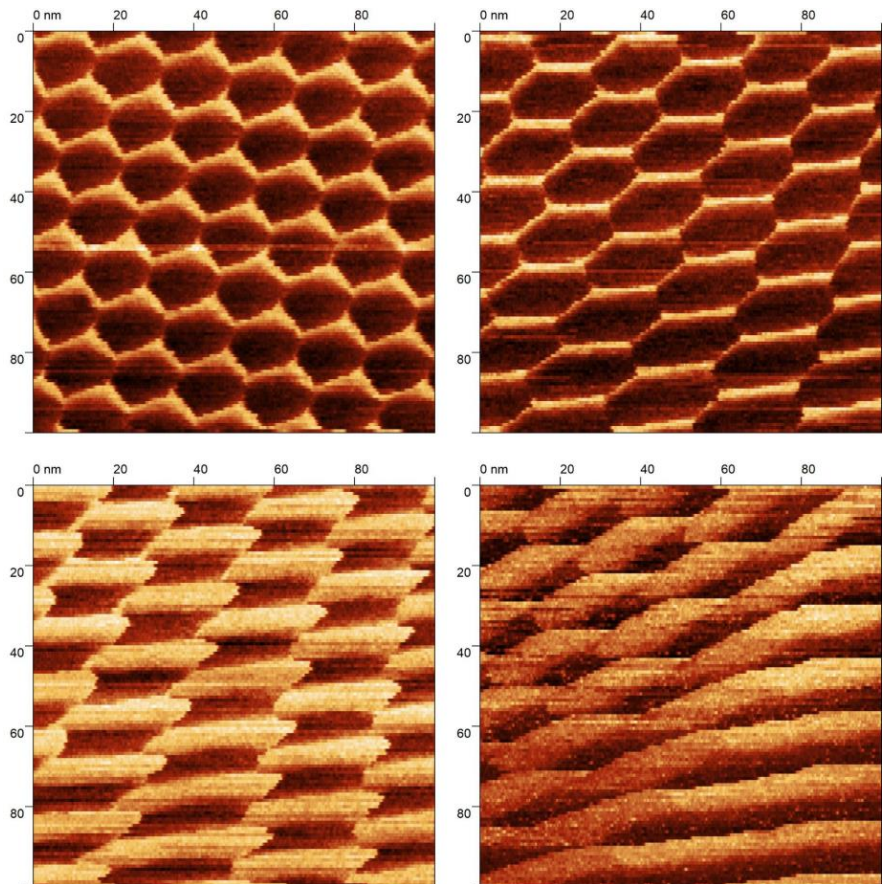

**Figure S9: Extreme moire heterostrain in Device A**
CAFM moire images of Device A, our most strained sample. Note it has a high degree of strain inhomogeneity, as evidenced by the varying moire size within the right and bottom pairs of images.

# Superlattice Dirac peak temperature dependence

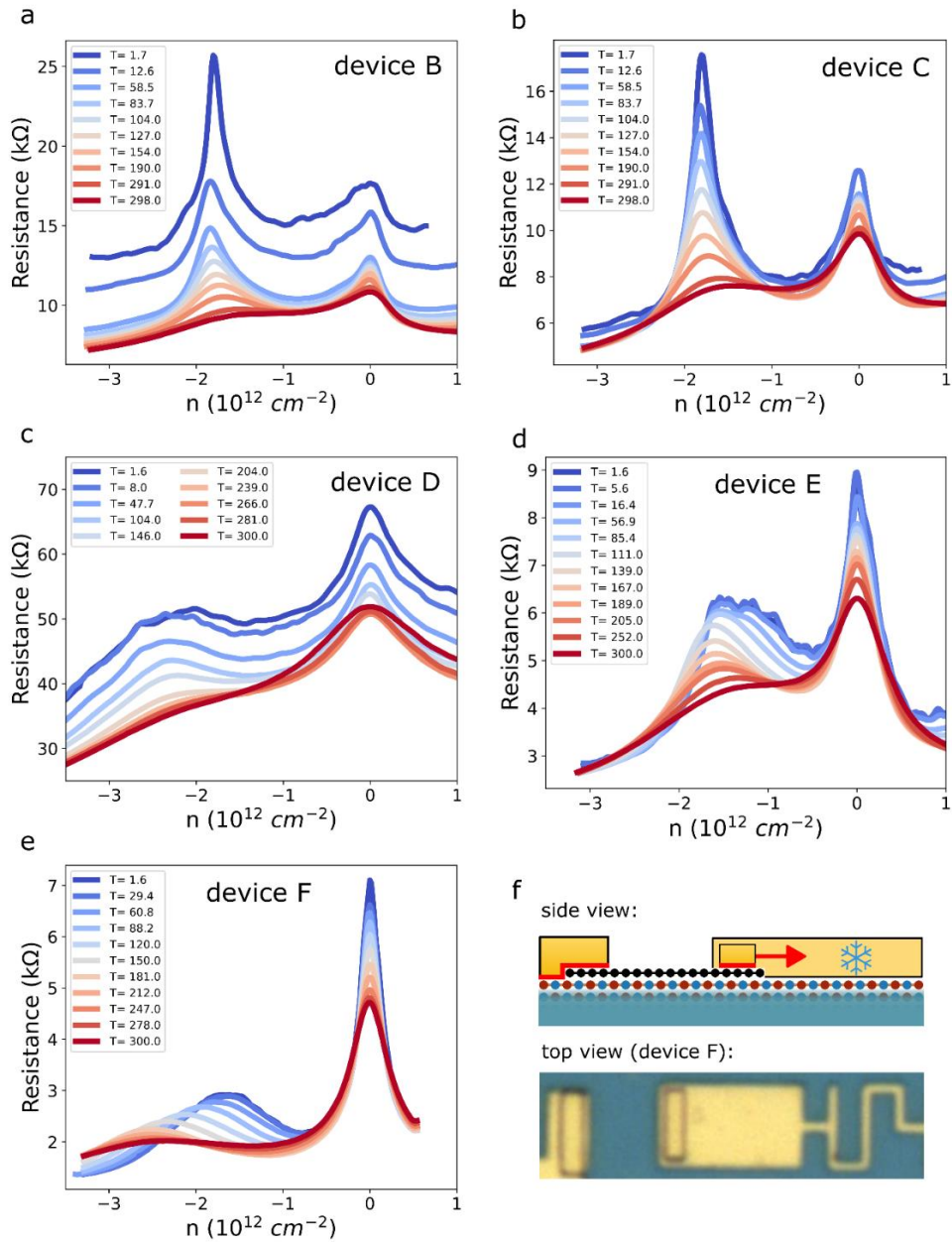

**Figure S10: Temperature dependent gate sweeps for various devices**
**a-e:** Temperature dependent gate sweeps for 5 separate devices. Red to blue color scale corresponds to temperatures from 300K to 1.6K.
**f:** Side view schematic indicating thermal contraction of the anchor with temperature and top view optical image of device F.

| Device | Anchor (um) | Channel (um) | ratio |
|--------|-------------|--------------|-------|
| B | 1.5 | 3 | 0.5 |
| C | 1.5 | 4 | 0.375 |
| D | 1.5 | 4.5 | 0.33 |
| E | 1.3 | 2.7 | 0.48 |
| F | 5.2 | 3.2 | 1.625 |

**Table 2: Table of the device dimension presented in Figure S10.**

     Upon cooling, the superlattice peak for most of our devices shifts outward, to higher densities (Figure S10a-d). This effect is also observed in twisted bilayer graphene[13,14] and is discussed in the main text. However, for one device, the dependence is opposite, exhibiting an overall shift to lower densities and larger superlattice unit cells as the sample is cooled to 1.6K. This suggests that the graphene is being stretched as it is cooled down. We attribute this effect to thermal contraction of the anchoring drain electrode, which should have a large effect for this device as the metal anchor is ~1.75x the length of the graphene channel (Figure S10f). The anchor is held only by its friction with the hBN, so we expect it to contract inward equally on all sides. From 293K to 4K the relative contraction of gold is ~0.3%[15], so the graphene should receive an additional strain of 0.3% *1.75/2 = ~0.28%. This is a substantial amount of thermally-induced strain. By contrast, the other devices that do not exhibit thermal contraction effects have a metal anchor that is only half the length of the graphene channel, which would result in a 3.5x smaller thermally-induced strain (0.075%).

     In addition, variations in the O2 plasma etch performed on the electrodes lead to different levels of friction with the graphene. In some cases, this results in slipping and a lack of one-to-one motion during manipulations, as was the case for the devices B and C presented in Figure 3 of the main text. Consequently, significantly larger motions (~100x) of the handle were necessary to impart strain. This provides further reasoning as to why for these devices we do not observe a decrease in carrier densities due to the thermal contraction of the anchors upon cooling.

# Additional Magnetotransport data

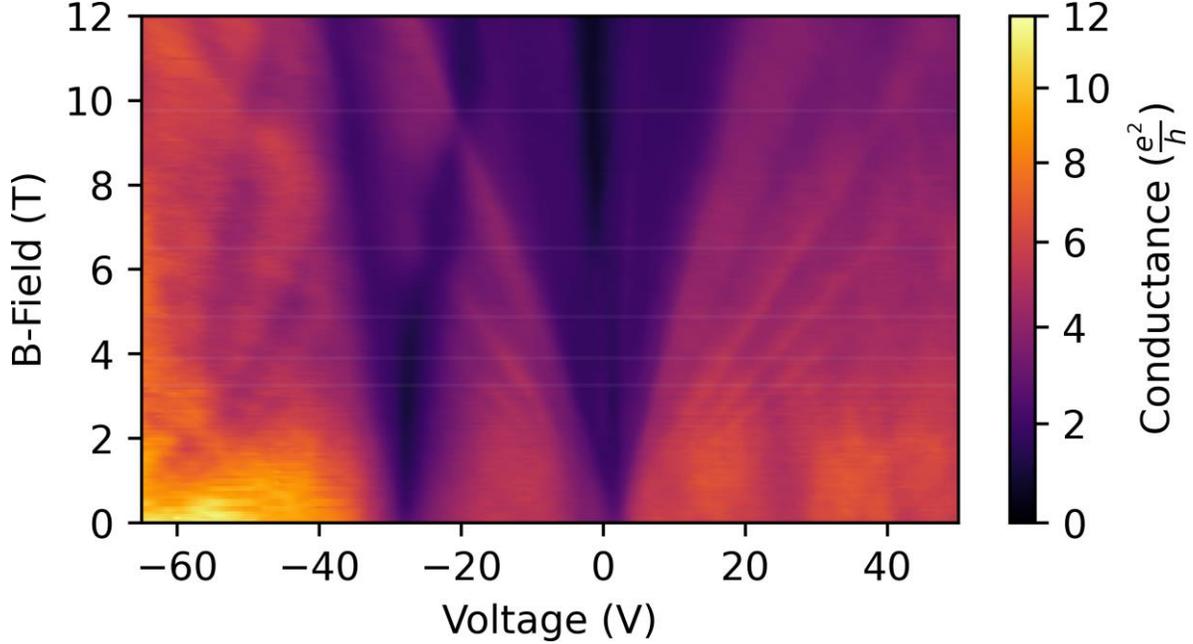

**Figure S11: 1.6K magnetotransport of device B.** Before cooling down, the sample had an average: uniaxial strain of 0.35±0.05%, twist angle of 0.01±0.05°, and strain angle of 14±8° with respect to the graphene lattice vector where the uncertainties are the standard deviation from the CAFM dataset. A constant resistance value of 10.54 kΩ is subtracted to align the $\nu$=2 plateau at $2e^2/h$.

## Theoretical analysis of strained graphene-hBN

In our theoretical calculation, we assume that only graphene has a uniform strain, characterized by the strain strength $\epsilon$ and strain angle $\phi$. We set $\phi = 0$ to align with the zigzag direction of moire hexagonal pattern of g-hBN. The moire pattern of strained g-hBN is defined by $\mathbf{G}_j = \mathbf{b}_j - \mathbf{b}'_j$, where $\mathbf{b}_j$, $\mathbf{b}'_j$ and $\mathbf{G}_j$ are the reciprocal lattice vector of the strained graphene, hBN, and moire lattice respectively. The moire real lattice vector is given by $\mathbf{G}_i \cdot \mathbf{L}_j = 2\pi \delta_{ij}$.

To investigate the relaced lattice structure of the strained g-hBN, we apply the continuum method[16,17]. For the analysis of the electronic properties, we us the continuum Hamiltonian with lattice relaxation[18,19].

### Moire Pattern

In this section, we define the moire pattern for strained graphene-hBN. We consider the bilayer system constructed by monolayer graphene and hBN. We take the xy-plane in the layers and the z-axis along their perpendicular direction. The lattice constants of graphene and hBN are $a = 0.246$ nm and $a_{hBN} = 0.2504$ nm respectively.

Before we derive moire lattice of strained graphene-hBN, we briefly review moire pattern the non-strained case. We consider that hBN is rotated by $\theta$ relative to graphene. The lattice vectors of graphene are defined as $\mathbf{a}_1^0 = a(1,0)$ and $\mathbf{a}_2^0 = a\left(\frac{1}{2}, \frac{\sqrt{3}}{2}\right)$, and for hBN, it is written as $\mathbf{a}'_j = R(\theta)\left(\frac{a_{hBN}}{a}\right)\mathbf{a}_j^0$ for $j = 1,2$ where $R(\theta)$ is the 2d rotation matrix. The reciprocal lattice vector are given by $\mathbf{b}_1^0 = \left(\frac{4\pi}{\sqrt{3}a}\right)\left(\frac{\sqrt{3}}{2}, -\frac{1}{2}\right)$ and $\mathbf{b}_2^0 = \left(\frac{4\pi}{\sqrt{3}a}\right)(0,1)$ for graphene, and $\mathbf{b}'_j = R(\theta)\left(\frac{a}{a_{hBN}}\right)\mathbf{b}_j^0$ for hBN. The corners of the graphene and hBN Brillouin zones are written as $\mathbf{K}_\xi = \frac{\xi(2\mathbf{b}_1^0 + \mathbf{b}_2^0)}{3}$ and $\mathbf{K}'_\xi = R(\theta)\left(\frac{a}{a_{hBN}}\right)\mathbf{K}_\xi$ respectively.

While the system generally has a commensurate moire period at the specific twist angle, for an angle small enough, we can define the incommensurate moire period by $\mathbf{G}_j^0 = \mathbf{b}_j^0 - \mathbf{b}'_j$ for $j = 1,2$. The moire lattice vector $\mathbf{L}_j$ is given from $\mathbf{G}_i^0 \cdot \mathbf{L}_j^0 = 2\pi\delta_{ij}$. The angle between $\mathbf{L}_1^0$ and x-axis is

$$\phi_M = \arctan\left(\frac{-\sin\theta}{1 + \delta - \cos\theta}\right), \quad (A5)$$

where $\delta = \frac{(a_{hBN} - a)}{a}$.

Next, we move to the case of strained graphene-hBN. In this model, we ignore the strain on hBN because in our system strain is only applied to the graphene. The primitive lattice vector of strained monolayer graphene is given by $\mathbf{a}_j = (1 + \mathcal{E})\mathbf{a}_j^0$, and the reciprocal lattice vector is $\mathbf{b}_j = (1 + \mathcal{E})^{-1,T}\mathbf{b}_j^0$, where

$$\mathcal{E} = R\begin{pmatrix} \epsilon & 0 \\ 0 & -\nu_P\epsilon \end{pmatrix} R^T \quad (A6)$$

$$R = \begin{pmatrix} \cos(\phi + \phi_M) & -\sin(\phi + \phi_M) \\ \sin(\phi + \phi_M) & \cos(\phi + \phi_M) \end{pmatrix}$$

is the strain matrix. $\epsilon$ and $\phi$ are the strength and direction of the strain. $\nu_P = 0.16$ is the Poisson's ratio of monolayer graphene[11,12]. We set $\phi = 0$ to align to $L_1^0$ at each twist angle. $\phi = 60°$ gives the same moire pattern of $\phi = 0°$ due to $C_{2z}$ symmetry of the strain matrix (A6) and $C_{3z}$ symmetry of non-strained graphene-hBN. Therefore we only consider $(0° \leq \phi < 60°)$ in this model.

The moire reciprocal lattice vector for the strained Graphene-hBN can be defined by

$$\begin{aligned} \mathbf{G}_j &= \mathbf{b}_j - \mathbf{b}'_j \\ &= \left[(1 + \mathcal{E})^{-1,T} - \left(\frac{a}{a_{hBN}}\right)R(\theta)\right]\mathbf{b}_j \end{aligned} \quad (A7)$$

for each strain parameter $\epsilon$ and $\phi$. The moire lattice vector is given by $\mathbf{G}_i \cdot \mathbf{L}_j = 2\pi \delta_{ij}$.

## Continuum method for lattice relaxation

For each strain parameter $\epsilon$ and $\phi$, we calculate the optimized graphene and hBN lattice structure by using the continuum method[16,17] while keeping the moire unit cell strained.

We define the displacement vector of each atom as $\mathbf{u}(R_X^{(l)})$, where $\mathbf{R}_X^{(l)}$ is the atom position of the sublattice $X(=A,B)$ in layer $l$. When moire period is long enough relative to the atomic length scale, the displacement vector $\mathbf{u}$ smoothly varies in the moire length scale. As a result, we can adapt the continuum approximation for the displacement vector and we can replace it as the continuum function of real space $\mathbf{u}(R_X^{(l)}) \rightarrow \mathbf{u}^{(l)}(\mathbf{r})$.

We introduce the total lattice energy as the function of the displacement vector $U = U[\mathbf{u}^{(1)}(\mathbf{r}), \mathbf{u}^{(2)}(\mathbf{r})] = U_B + U_E$, and the optimized lattice structure is given by solving $\frac{\partial U}{\partial u_\mu^{(l)}} = 0$. $U_B$ is the binding energy between layers

$$U_B = \int d^2\mathbf{r} \sum_{j=1}^{3} 2V_0 \cos[\mathbf{G}_j \cdot \mathbf{r} + \mathbf{b}_j^0 \cdot (\mathbf{u}^{(2)} - \mathbf{u}^{(1)}) + \psi_0], \quad (A8)$$

where $\mathbf{b}_3^0 = -\mathbf{b}_1^0 - \mathbf{b}_2^0$, $\mathbf{G}_3 = -\mathbf{G}_1 - \mathbf{G}_2$. We take $V_0 = 0.202$ eV/nm$^2$ and $\psi_0 = -0.956$[17]. We ignore the constant term of the binding energy because it disappears in the equation $\partial U / \partial u_\mu^{(l)} = 0$.

The elastic energy $U_E$ is written in a standard form[20,21] as

$$U_E = \sum_{l=1}^{2} \frac{1}{2} \int \left[ (\mu^{(l)} + \lambda^{(l)})(u_{xx}^{(l)} + u_{yy}^{(l)})^2 \right.$$
$$\left. + \mu^{(l)} \left\{ (u_{xx}^{(l)} - u_{yy}^{(l)})^2 + 4(u_{xy}^{(l)})^2 \right\} \right] d^2\mathbf{r}, \quad (A9)$$

where $\lambda^{(1)} = 3.25$ eV/A$^2$ and $\mu^{(1)} = 9.57$ eV/A$^2$ are graphene's Lame factors, and $\lambda^{(2)} = 3.5$ eV/A$^2$ and $\mu^{(2)} = 7.8$ eV/A$^2$ are for hBN[7,22,23]. $u_{ij}^{(l)} = (\partial_i u_j^{(l)} + \partial_j u_i^{(l)})/2$ is the strain tensor. Here we ignore the elastic energy caused by the strain matrix because it only gives a constant term due to the uniformity of strain.

To solve the equation $\partial U / \partial u^{(l)\mu} = 0$, we define the Fourier components of the displacement vector as

$$\mathbf{u}(\mathbf{r}) = \sum_{\mathbf{G}} \mathbf{u}_{\mathbf{G}}\, e^{i\mathbf{G}\cdot\mathbf{r}}. \qquad (A10)$$

where $\mathbf{G} = m_1\mathbf{G}_1 + m_2\mathbf{G}_2$ are the reciprocal lattice vector of strained moire. We also introduce $f_{\mathbf{G},j}$ by

$$\sin\left[\mathbf{G}_j\cdot\mathbf{r} + \mathbf{b}_j^0\cdot\left(\mathbf{u}^{(2)}(\mathbf{r}) + \mathbf{u}^{(2)}(\mathbf{r})\right)\right] = \sum_{\mathbf{G}} f_{\mathbf{G},j}\, e^{i\mathbf{G}\cdot\mathbf{r}}. \qquad (A11)$$

After some calculation using the above Fourier components, $\partial U / \partial u_\mu^{(l)} = 0$ for $(\mu = x, y)$ and $(l = 1, 2)$ gives the self-consistent equation as follows,

$$\begin{aligned}
\mathbf{u}_{\mathbf{G}}^{(1)} &= -2V_0 \sum_{j=1}^{3} f_{\mathbf{G},j}\left[\widehat{K}_{\mathbf{G}}^{(1)}\right]^{-1} \mathbf{b}_j^0 \\
\mathbf{u}_{\mathbf{G}}^{(2)} &= +2V_0 \sum_{j=1}^{3} f_{\mathbf{G},j}\left[\widehat{K}_{\mathbf{G}}^{(2)}\right]^{-1} \mathbf{b}_j^0,
\end{aligned} \qquad (A12)$$

where

$$\widehat{K}_{\mathbf{G}}^{(l)} = \begin{pmatrix} (\lambda^{(l)} + 2\mu^{(l)})G_x^2 + \mu^{(l)}G_y^2 & (\lambda^{(l)} + \mu^{(l)})G_x G_y \\ (\lambda^{(l)} + \mu^{(l)})G_x G_y & (\lambda^{(l)} + 2\mu^{(l)})G_y^2 + \mu^{(l)}G_x^2 \end{pmatrix}.$$

By using Eq. A10, A11, and A12, we numerically solve the self-consistent equation while keeping the strained moire period.

## Continuum Hamiltonian

Here we define the continuum Hamiltonian for strained graphene-hBN following[18,19]. The Hamiltonian for valley $\xi$ is written as

$$H^{(\xi)}(\mathbf{k}) = \begin{pmatrix} H_G(\mathbf{k}) & U^\dagger \\ U & H_{hBN} \end{pmatrix},$$

where $H_G(\mathbf{k})$ and $H_{hBN}$ are the $2\times 2$ Hamiltonian of distorted monolayer graphene and hBN. The $H_G(\mathbf{k})$ is given by

$$H_G(\mathbf{k}) = -\hbar v\left[(1+\mathcal{E})^{-1}\left(\mathbf{k} + \frac{e}{\hbar}\mathbf{A}\right)\right]\cdot\boldsymbol{\sigma},$$

where $v$ is the graphene's band velocity, $\boldsymbol{\sigma} = (\xi\sigma_x, \sigma_y)$ and $\sigma_x$, $\sigma_y$ are the Pauli matrices in the sublattice space $(A, B)$.

We take $\hbar v/a = 2.14$ eV[24]. The $\mathbf{A}$ is the strain-induced vector potential that is given by[20,25,26]

$$\mathbf{A}^{(l)} = \xi \frac{3}{4} \frac{\beta \gamma_0}{ev} \begin{pmatrix} \epsilon_{xx} - \epsilon_{yy} \\ -2\epsilon_{xy} \end{pmatrix}.$$

$\gamma_0 = 2.7$ eV is the nearest neighbor transfer energy of intrinsic graphene and $\beta \approx 3.14$. The effective Hamiltonian of hBN $H_{hBN}$ is given by

$$H_{hBN} = \begin{pmatrix} V_N & 0 \\ 0 & V_B \end{pmatrix},$$

where $V_N$ and $V_B$ are the on-site potential of nitrogen and boron. We ignore the $\mathbf{k}$ dependence in this model. $U$ is the inter-layer coupling which written as

$$U = \sum_{j=1}^{3} U_j \, e^{i\xi \left[ \mathbf{q}_j \cdot \mathbf{r} + \mathbf{Q}_j \cdot \left( \mathbf{u}^{(2)}(\mathbf{r}) - \mathbf{u}^{(1)}(\mathbf{r}) \right) \right]},$$

where we define

$$q_1 = K^{(1)} - K^{(2)}, \quad q_2 = q_1 + \mathbf{G}_1, \quad q_3 = q_1 + \mathbf{G}_1 + \mathbf{G}_2$$
$$Q_1 = K_+, \quad Q_2 = Q_1 + \mathbf{b}_1, \quad Q_3 = Q_1 + \mathbf{b}_1 + \mathbf{b}_2.$$

and

$$U_1 = \begin{pmatrix} 1 & 1 \\ 1 & 1 \end{pmatrix}, \quad U_2 = \begin{pmatrix} 1 & \omega^{-\xi} \\ \omega^{+\xi} & 1 \end{pmatrix}, \quad U_3 = \begin{pmatrix} 1 & \omega^{+\xi} \\ \omega^{-\xi} & 1 \end{pmatrix}.$$

The parameters $t_0 = 152$ meV is the interlayer coupling strength[27].

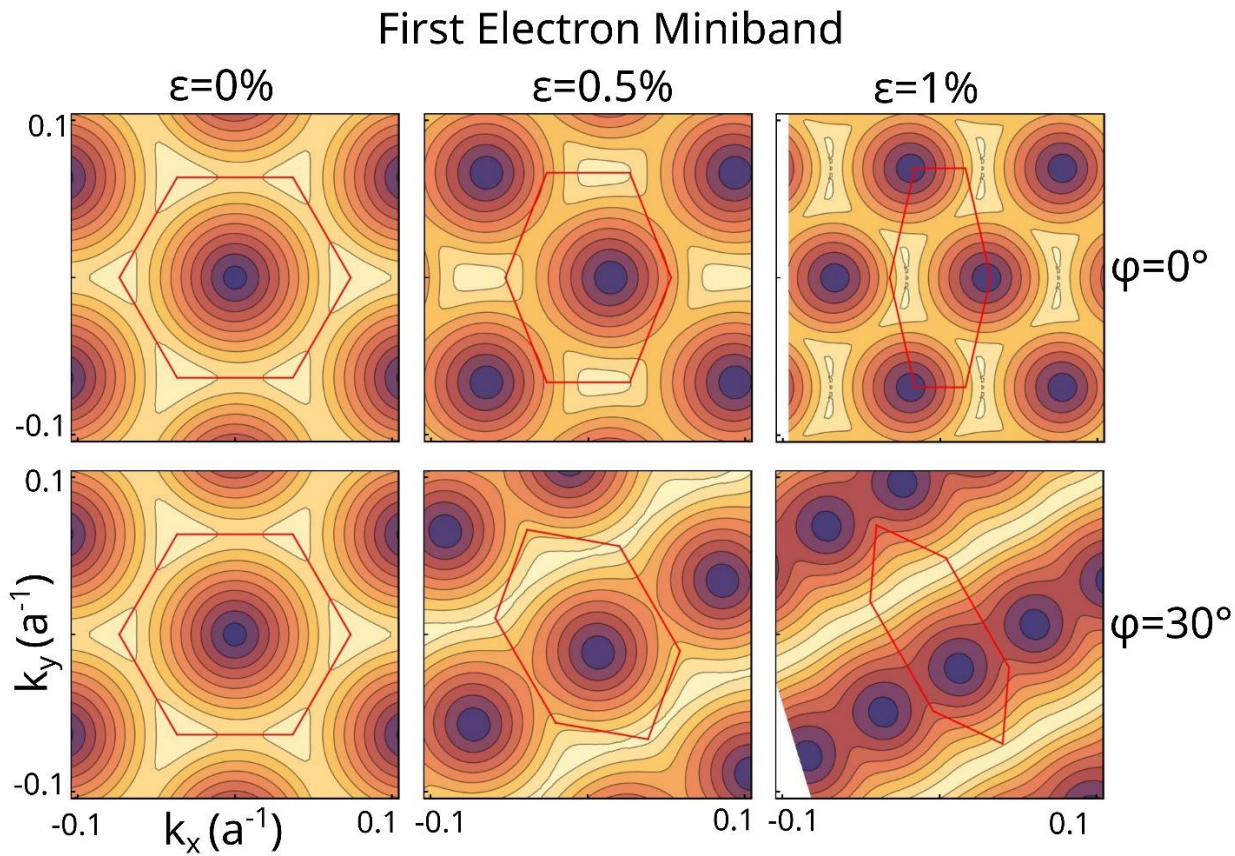

**Figure S12: Contour plots of first electron mini band**
Contour intervals are 10 meV.

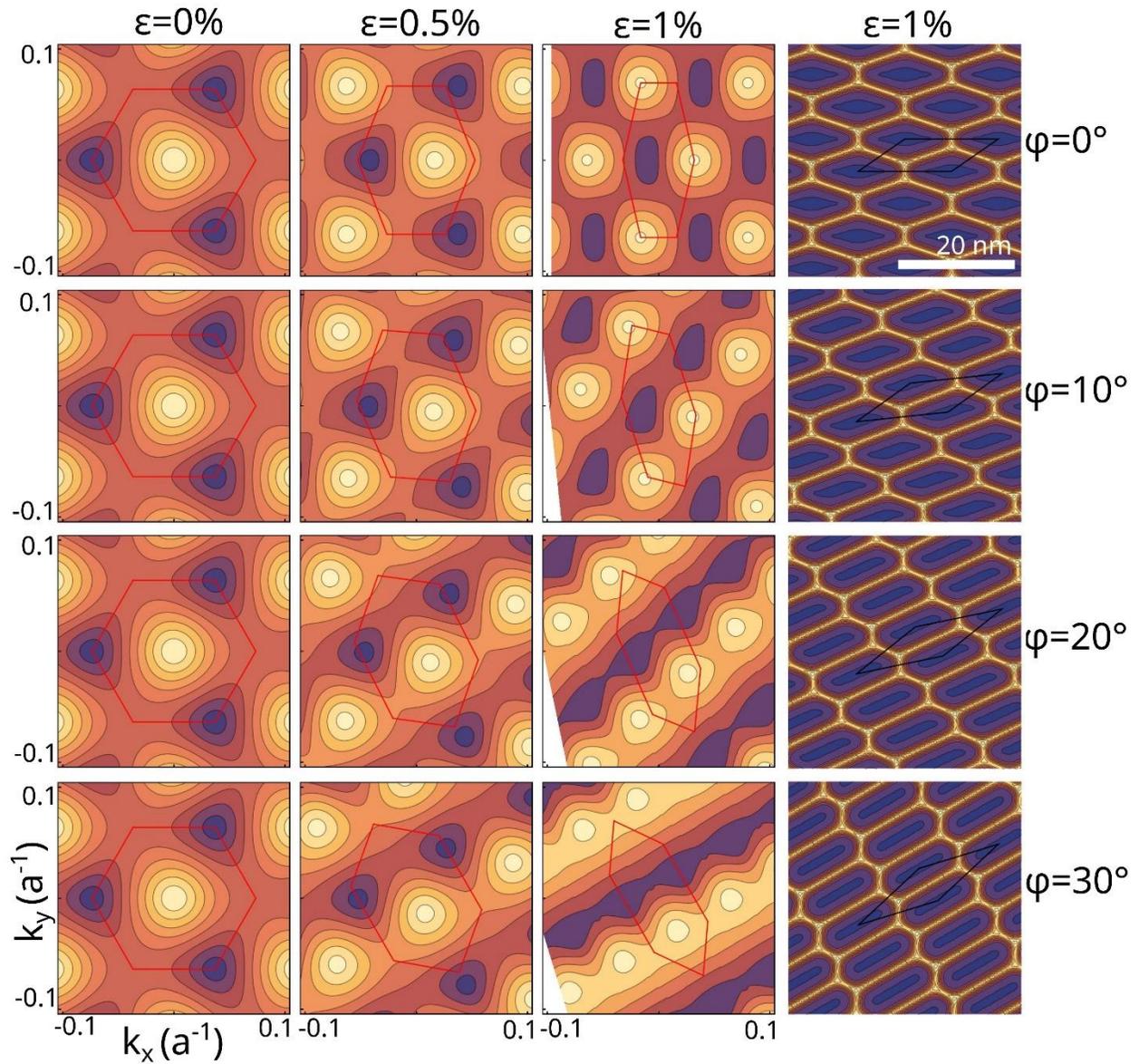

**Figure S13: First hole mini band contours dependence on strain angle**
Band structure contour intervals are 10 meV. The right column shows the real space moire structure for 1% strain. Real space contour intervals are 0.2 eV/nm$^2$.

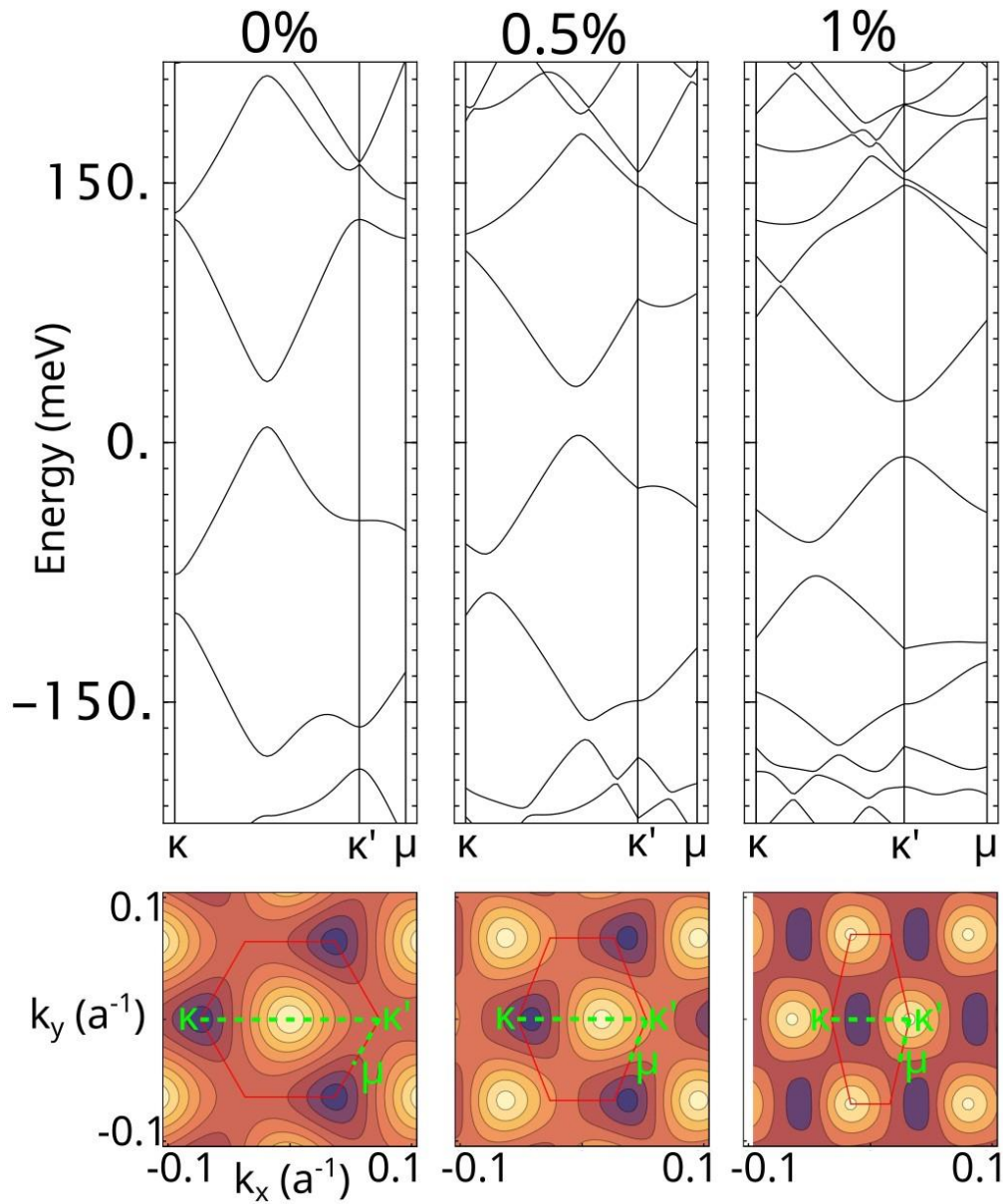

**Figure S14: Band structure cuts for several strain values at 0° strain angle**
Upper plots are band structure cuts following the green dashed line in the 2D contour plots below. Contour intervals are 10 meV.